 \documentclass[final,5p,times,twocolumn,authoryear]{elsarticle}

\usepackage{amssymb}
\usepackage{lipsum}
\usepackage{url}
\usepackage{amsmath}
\usepackage{tabularx}
\journal{Astronomy $\&$ Computing}

\begin{document}

\begin{frontmatter}

%% Title, authors and addresses

%% use the tnoteref command within \title for footnotes;
%% use the tnotetext command for theassociated footnote;
%% use the fnref command within \author or \affiliation for footnotes;
%% use the fntext command for theassociated footnote;
%% use the corref command within \author for corresponding author footnotes;
%% use the cortext command for theassociated footnote;
%% use the ead command for the email address,
%% and the form \ead[url] for the home page:
%% \title{Title\tnoteref{label1}}
%% \tnotetext[label1]{}
%% \author{Name\corref{cor1}\fnref{label2}}
%% \ead{email address}
%% \ead[url]{home page}
%% \fntext[label2]{}
%% \cortext[cor1]{}
%% \affiliation{organization={},
%%            addressline={}, 
%%            city={},
%%            postcode={}, 
%%            state={},
%%            country={}}
%% \fntext[label3]{}

\title{Determining Research Priorities Using Machine Learning}

%% use optional labels to link authors explicitly to addresses:
%% \author[label1,label2]{}
%% \affiliation[label1]{organization={},
%%             addressline={},
%%             city={},
%%             postcode={},
%%             state={},
%%             country={}}
%%
%% \affiliation[label2]{organization={},
%%             addressline={},
%%             city={},
%%             postcode={},
%%             state={},
%%             country={}}

% \author[first]{Author name}
% \affiliation[first]{organization={University of the Moon},%Department and Organization
%             addressline={}, 
%             city={Earth},
%             postcode={}, 
%             state={},
%             country={}}

\author[first]{Brian A. Thomas}
\author[second]{Harley Thronson}
\author[first]{Anthony Buonomo}
\author[third]{Louis Barbier}

\affiliation[first]{organization={Heliophysics Science Division, NASA Goddard Space Flight Center},%Department and Organization
            addressline={8800 Greenbelt Rd.}, 
            city={Greenbelt},
            postcode={MD 20771}, 
            country={USA}}

\affiliation[second]{organization={NASA (retired)}, 
            addressline={617 Tivoli Passage}, 
            city={Alexandria},
            postcode={VA 22314}, 
            country={USA}}

\affiliation[third]{organization={Office of the Chief Scientist, NASA Headquarters}, 
            addressline={300 E Street SW}, 
            city={Washington},
            postcode={DC 20546}, 
            country={USA}}

% \begin{abstract}
% %% Text of abstract
% Example abstract for the astronomy and computing journal. Here you provide a brief summary of the research and the results.
% \end{abstract}

\begin{abstract}
We summarize our exploratory investigation into whether Machine Learning (ML) techniques applied to publicly available professional text can substantially augment strategic planning for astronomy. We find that an approach based on Latent Dirichlet Allocation (LDA) using content drawn from astronomy journal papers can be used to infer high-priority research areas. While the LDA models are challenging to interpret, we find that they may be strongly associated with meaningful keywords and scientific papers which allow for human interpretation of the topic models. 

 Significant correlation is found between the results of applying these models to the previous decade of astronomical research (“1998$-$2010” corpus) and the contents of the science frontier panel report which contains high-priority research areas identified by the 2010 National Academies’ Astronomy and Astrophysics Decadal Survey (“DS2010” corpus). Significant correlations also exist between model results of the 1998$-$2010 corpus and the submitted whitepapers to the Decadal Survey (“whitepapers” corpus). Importantly, we derive predictive metrics based on these results which can provide leading indicators of which content modeled by the topic models will become highly cited in the future. Using these identified metrics and the associations between papers and topic models it is possible to identify important papers for planners to consider. 

A preliminary version of our work was presented by \cite{thronson2021transforming} and \cite{thomas2022research}.
\end{abstract}

%%Graphical abstract
%\begin{graphicalabstract}
%\includegraphics{grabs}
%\end{graphicalabstract}

%%Research highlights
%\begin{highlights}
%\item Research highlight 1
%\item Research highlight 2
%\end{highlights}

% \begin{keyword}
% %% keywords here, in the form: keyword \sep keyword, up to a maximum of 6 keywords
% keyword 1 \sep keyword 2 \sep keyword 3 \sep keyword 4

% %% PACS codes here, in the form: \PACS code \sep code

% %% MSC codes here, in the form: \MSC code \sep code
% %% or \MSC[2008] code \sep code (2000 is the default)

% \end{keyword}

\begin{keyword}
Machine Learning \sep Strategic Planning \sep Astronomy Research
\end{keyword}

\end{frontmatter}

%\tableofcontents

%% \linenumbers

%% main text

\section{Introduction}
\label{introduction}

One of the most critical planning activities in the sciences is identifying credible priorities for investment. A highly regarded process of scientific prioritization is the National Academies’ Decadal Surveys. Among the principal challenges faced by this process is the Survey panelists’ necessity to assess a very large -- and rapidly growing -- amount of relevant information, specifically many tens of thousands of published research papers in journals (\cite{santamaria2018mining}). The potential input materials have thus increased greatly over the years in both variety and quantity, while the basic processes of the Surveys and other strategic planning activities have changed relatively little. The primary approach for the Surveys over the past half-century remains the same (\cite{dressler2015space}): a central steering committee of a couple dozen members supported by large specialty panels. This leads to the primary motivation of our work: How do we substantially improve the current very labor-intensive process of identifying the highest-priority science without adding additional personnel?

Advances in Artificial Intelligence (AI) over the past decade have been impressive. Increasingly powerful AI techniques have been developed which can comb through large corpora of unstructured text to reveal insight into their contents. Much recent work on use of AI to augment the science discovery and prioritization process has been broadly philosophical on the nature of “discovery” (\cite{kitano2016artificial,clark2022decentering,khalili2021toward}, among many others). Although interesting, our goal was instead to avoid such questions as “the nature of discovery” and instead undertake an assessment of the empirical determination of science priorities. That is, assess the outcome rather than the process of science prioritization and discovery.

There have recently been a few examples relevant to the determination of science prioritization that are somewhat similar to that which we describe here. For example, \cite{zelnio2020shaping} reports the successful use of Machine Learning (ML) to evaluate research literature for promising technologies and \cite{krenn2020predicting} demonstrate a method used to predict future trends in quantum physics. \cite{tshitoyan2019unsupervised} demonstrate a successful application of Natural Language Processing (NLP) applied to many thousands of research publications to reveal undiscovered materials. \cite{shi2023surprising} employ a hypergraph embedding model to demonstrate that surprising scientific advances often arise across teams and disciplines, rather than within them.

In this paper we report on our exploratory research to understand whether we can use ML techniques to provide useful insight into how the field of astronomy research is evolving. Important questions that should be answered include:

\begin{itemize}
    \item Which research areas are the prominent ones?
    \item Which research areas are growing (or declining) most significantly?
    \item Can we independently determine from the literature the most important topics of astronomy research? and
    \item How may we quantify the performance of these models relative to human perception of which areas of research are important?
\end{itemize}
\parshape=0

In order to determine the validity of machine predictions, we conducted an experiment using machine predictions derived from information contained in the astronomy literature published from 1998 - 2010 ("1998 - 2010" corpus) and compared it with the science priorities found in the 2010 National Academies’ Astronomy and Astrophysics Decadal Survey science frontier panels (“DS2010” corpus). Our earlier experiments in this area have been previously described in \cite{thronson2021transforming} and \cite{thomas2022research}. In this paper we update and significantly expand upon the experiments conducted in that work.

\section*{Topic Modeling}

\begin{figure*}[t!] % use figure* instead of figure for all cols
\centering
%\includesvg[width=\textwidth]{topic_modeling_pipeline.svg}
\includegraphics[width=.8\linewidth]{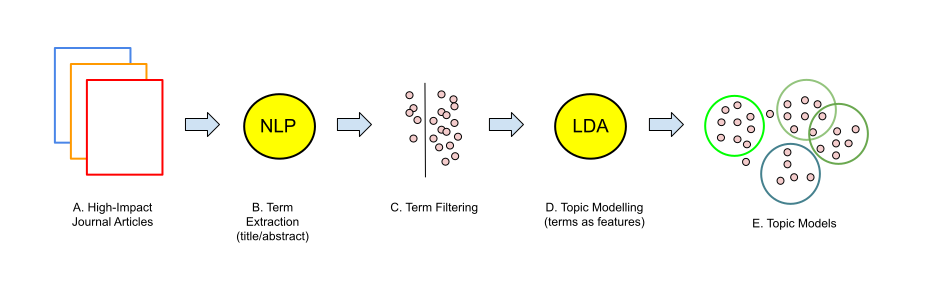}
\caption{Topic Modeling Pipeline. The diagram indicates how we create our topic models using refereed journal articles from high-impact journals (Appendix A). Steps A through E indicate key processing points which are described in the text.}
\label{fig:topic_model_pipeline}
\end{figure*}

A first key objective is to automate the classification of research areas. For this paper, we have investigated a classification approach to model topics of research in the literature using Latent Dirichlet Allocation (LDA, \cite{blei2003latent}), which is an unsupervised algorithm able to identify topics found in a large text corpus. A key challenge is the preparation of the input material so that the resulting LDA topic models align with areas of scientific research. The pipeline to determine topic models is shown in Figure \ref{fig:topic_model_pipeline} and starts with the formation of a corpus (step A). This corpus is created by downloading the titles and abstracts of published papers discovered using the Astrophysical Data Service (ADS,  \cite{kurtz2000astron}). In all cases papers were limited to those published in ‘high-impact’ astronomy journals in the decade 1998 - 2010 (see Appendix A for a description of the method used to determine high-impact astronomy journals).

We limited our analysis to the titles and abstracts because both concentrate on summarizing or detailing the research and related priority conclusions, as opposed to the other sections of a paper that may include topics of little relevance to our work: e.g., background or history, details of a program of observation, names of institutions or observational platforms, etc. Papers with abstracts of fewer than 100 characters were considered to have insufficient substantive content and were dropped. This filtering left approximately 77,000 papers for this time period.

In the next step (B), we use the SingleRank algorithm, an adaptation of the TextRank framework (\cite{mihalcea2004textrank}), to extract semantically significant terms from these titles and abstracts. These terms may be single or compound words and are in all cases “nouns”. Examples include “massive black hole”, “gravitational wave”, “star formation”, “agn”, and so on. In particular, we use the textacy (\cite{dewilde2020textacy}) library’s SingleRank implementation with a ten-word contextual window, count-based edge-weighting, and no positional bias. To help improve our key term extraction, particularly for the specialized vocabulary of scientific literature, we utilize the SciSpacy language model (\cite{neumann2019scispacy}) for normalization via lemmatization. This combination of SingleRank and SciSpacy facilitates the extraction of highly informative key terms.

After utilizing this SciSpacy-powered SingleRank algorithm for key term extraction on each document, we are left with a very large collection of terms. In order to winnow the terms down to those which are most informative, we apply several filters in step C. In particular, we remove any terms which occur in more than 20\% of documents, terms which occur fewer than 300 times total, and terms which have an average SingleRank score of less than 0.015. These filters are intended to remove terms which are overly general or which have a weak signal in the overall corpus.

Finally, subject matter experts further filter these terms by manually removing those terms which are unlikely to be unique to a particular type of research. These manually removed terms include those which refer to a common scientific technique (e.g., “photometry”), those which refer to a common type of data (e.g. “light curve”), ambiguous common terms (e.g., “new result”) or terms which refer to facilities/origins that may be related to many types of research (e.g., “Hubble Space Telescope”, “NASA”). Examples of resulting terms that we keep after filtering include “massive black hole”, “gravitational wave”, “star formation”, and significant acronyms such as “agn” (i.e., active galactic nucleus). There were a total of 1569 “blacklisted” terms which we identified and removed, leaving 399 terms for topic modeling.

The filtering of these terms better helps the modeling of the corpus in step D to create topics found in astrophysical research. Topic modeling uses the filtered term list as input “features” for the LDA. The resulting topics and associated models, step E, are associated with a (potentially large) collection of strongly associated terms (defined below and shown in the diagram as terms which appear inside the colored circles). The graphic in Figure \ref{fig:topic_model_pipeline} only shows a few terms in each topic for illustration, in particular to draw attention to the fact that sometimes strongly associated terms may be shared by two or more topics.

The optimal number of LDA topics (in our case, 125) was determined using a maximum coherence method, an approach that was systematized by \cite{roder2015exploring} and utilizes the UMass coherence measure (\cite{mimno2011optimizing}). This methodology builds on foundational work by \cite{newman2010automatic}, who first explicitly introduced the task of evaluating topic coherence, as a type of measure which seeks to align with qualitative human judgements of coherence. Irrespective of the number of topics chosen – we examined cases with no fewer than 25 topics – we always obtain a very small minority of topics that are an apparent mix of terms that make no obvious sense in that they collect documents with seemingly random criteria. The documents strongly associated with these types of topics tend to be those few non-scientific or broadly descriptive papers that are hard to associate with any specific type of research and so are problematic to model by the algorithm. This is an inescapable outcome for the LDA modeling of any large corpus drawn from an assortment of published material.

Latent Dirichlet Allocation (LDA) models a corpus as a collection of topics, where each topic is a distribution over terms, and each document is a mixture of these topics. Upon applying LDA, each document (\( d \)) is associated with a set of topics (\( t \)) through inference scores (\( I(t, d) \)). These scores represent the document's distribution over topics, indicating the strength of association between terms in the given document and each topic. A document which is uniquely connected to a topic would have an inference score of 1.0, whereas a document with no relationship to a topic will have a score of 0.0. In practice, no document is uniquely associated with any single topic, but is a blend of several topics. Many documents are “strongly associated” with a singular topic, which we define as 0.5 or greater topic inference for that paper. On the other hand, some documents are “poorly associated” and have almost uniformly low inference with every topic; that is, essentially an inference score of approximately 1/(number of topics). There are many documents which lie between these two extremes and have a peak inference which is between 0.3 - 0.4 for each of two or three topics associated with a document.

Understanding the nature of each topic is important in order assess its relationship to research. Human interpretation may be achieved using both terms and documents which are strongly associated to the topic. We have already defined above criteria for determination of strongly associated documents but strongly associated term criteria require some further explanation and definition as follows. Because topics are distributions over terms, all of the thousands of terms are at some level associated with each topic. Moreover, some terms may be ranked highly for a given topic, but also ranked highly for many other topics due to these terms’ frequency across the corpus. Consequently, just looking at the most probable terms for a given topic may fail to provide easily interpretable topics and uniquely characterize a topic. To remedy this issue, we use a “term relevance” metric \cite{sievert2014ldavis} to intelligently re$-$weight a given topic’s terms to determine which terms are most “strongly associated,” defined in this work as those terms that have a relevance log probability greater or equal than -2.5. Examples of labels we may assign to topics include “massive black hole evolution,” “dwarf galaxy evolution,” “early dark-energy universe modeling,” and “high-redshift galaxy.”

The stochastic nature of LDA should be mentioned as it could impact the interpretation of the results. Because the LDA model defines posteriors which are computationally intractable, LDA is often trained with variational inference, which requires initializations for topic-word and document-topic matrices. The initializations are randomly drawn from gamma distributions and they directly affect the approximate posteriors, the trained model. Consequently, each time the algorithm is run on the same input, but with a differing random seed, the output topic models will vary. To understand this variation we examined how much a topic appeared in repeated runs. We recreated the topic models for ten runs, each run initialized with a differing random seed, and used strongly associated topic terms of each topic to construct Cosine Similarity Vectors (CSVs) between every topic in every run. The magnitude of the term CSV used the estimated number of occurrences of the term in documents associated with the topic (\cite{mabey2021pyldavis}) in a given run. The term CSV for each topic were then compared with all other topic term CSV in all other runs and the maximum value of cosine similarity was kept between the examined topic and the best matching topic in another run. This collection of nine measured maximum values for cosine similarity was then averaged to find the mean best cosine similarity of the topic.

\begin{figure}[t!]
\centering
\includegraphics[width=.8\linewidth]{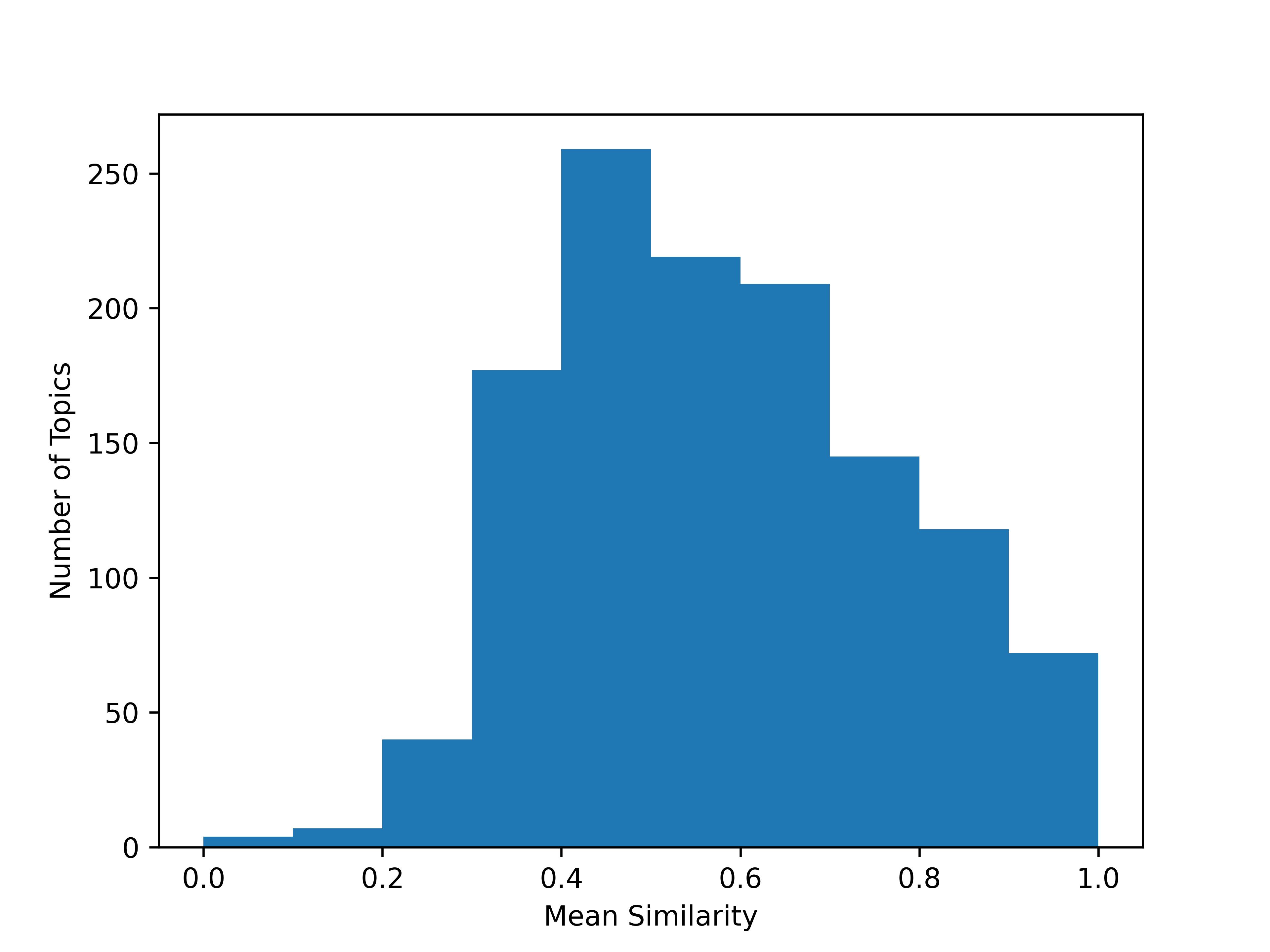}
\caption{Gauging stability of derived topic models. The diagram indicates how interrelated topic models are between separate runs of LDA (trained using the 1998$-$2010 corpus) using different random seeds. Ten runs generating topic models were created and topics in each run then cross-compared to topics in other runs using cosine similarity based on the strongly associated term for each topic. “Mean Similarity,” a gauge of topic stability, was calculated in two steps. First by finding the best (highest) cosine similarity for a given topic to all other topics in another run. This was then repeated for all runs to yield nine measurements of cosine similarity for the topic which were then averaged to yield the mean similarity.}
\label{fig:topic_stability}
\end{figure}

Results of this assessment of similarity, using papers from the 1998$-$2000 corpus,  appear in Figure \ref{fig:topic_stability}, where we find that more than half of the topics ($\approx59\%$) for all runs had a mean cosine similarity of greater than 50\% indicating significant similarity exists between a given topic in one run and some best-matching topic in another. In other words, the algorithm is creating similar topics each time it is run with a different random seed for most of the topics. This indicates the majority of the derived topics were approximately stable in occurrence from run to run, although very few topics are so stable ($\lessapprox 5\%$ of all topics) that they occur with little change from run to run ($\gtrapprox 90\%$ mean similarity).

\section*{Analysis}

We have developed several metrics designed to help answer key questions identified in Section 1. The first metric is the “Topic Contribution Score” (TCS) which determines the overall representation of a given topic in a corpus and which we use to infer how much the population of researchers is studying (publishing) work related to a given research topic \( t \) (Question 1). TCS(\( t \)) may be calculated as
\begin{equation}
    \text{TCS}(t) = \sum_{d=D_1}^{d=D_n} I(t, d),
\end{equation}
where \( I(t, d) \) is the LDA inference score or inferred amount of how much topic \( t \) is represented in document \( d \) in a collection of documents of size n. \( I(t, d) \) has a value between 0 and 1 and the sum of all \( I(t, d) \) for all topics always sums to 1 for any document \( d \). The TCS for the body of astronomical literature (TCS\(_{1998-2010}\)) may therefore be calculated by obtaining the inference scores for the selected content of each paper in the corpus (I\(_{1998-2010}\)) by applying the derived LDA topic models on a paper-by-paper basis (paper abstract and title content serve as the documents).

\begin{figure*}[t!]
\centering
%\includesvg[width=\textwidth]{topic_contribution_score_pipeline.svg}
\includegraphics[width=.8\linewidth]{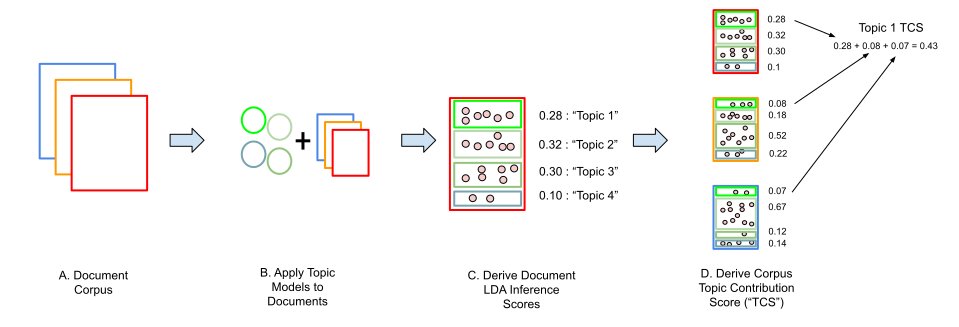}
\caption{Topic Contribution Score Pipeline. This shows the pipeline used to calculate “Topic Contribution Score” (TCS) or the total contribution of a topic to an example corpus of three documents and four topics. This pipeline is used for the determination of topic contributions for our various corpora which include the 1998$-$2010 corpus (documents are the journal papers abstracts and titles), the Decadal Survey (documents are the text blocks found in the science frontier panel chapters 1-4), and submitted whitepaper content (documents are the whole text of the papers). Detailed description of steps A through D appear in the text.}
\label{fig:topic_contribution_score_pipeline}
\end{figure*}

Figure \ref{fig:topic_contribution_score_pipeline} summarizes steps used to calculate TCS. First, a corpus is assembled (Step A) and then the derived LDA topic models applied to each document (Step B) to determine the contributions (inference values) of each topic in the given document (Step C). The topic contributions in each document are then summed, by each topic \( t \), across the corpus to create the TCS(\( t \)), or Topic Contribution Score of topic \( t \) (Step D).

\begin{figure*}[t!]
\centering
\includegraphics[width=\textwidth]{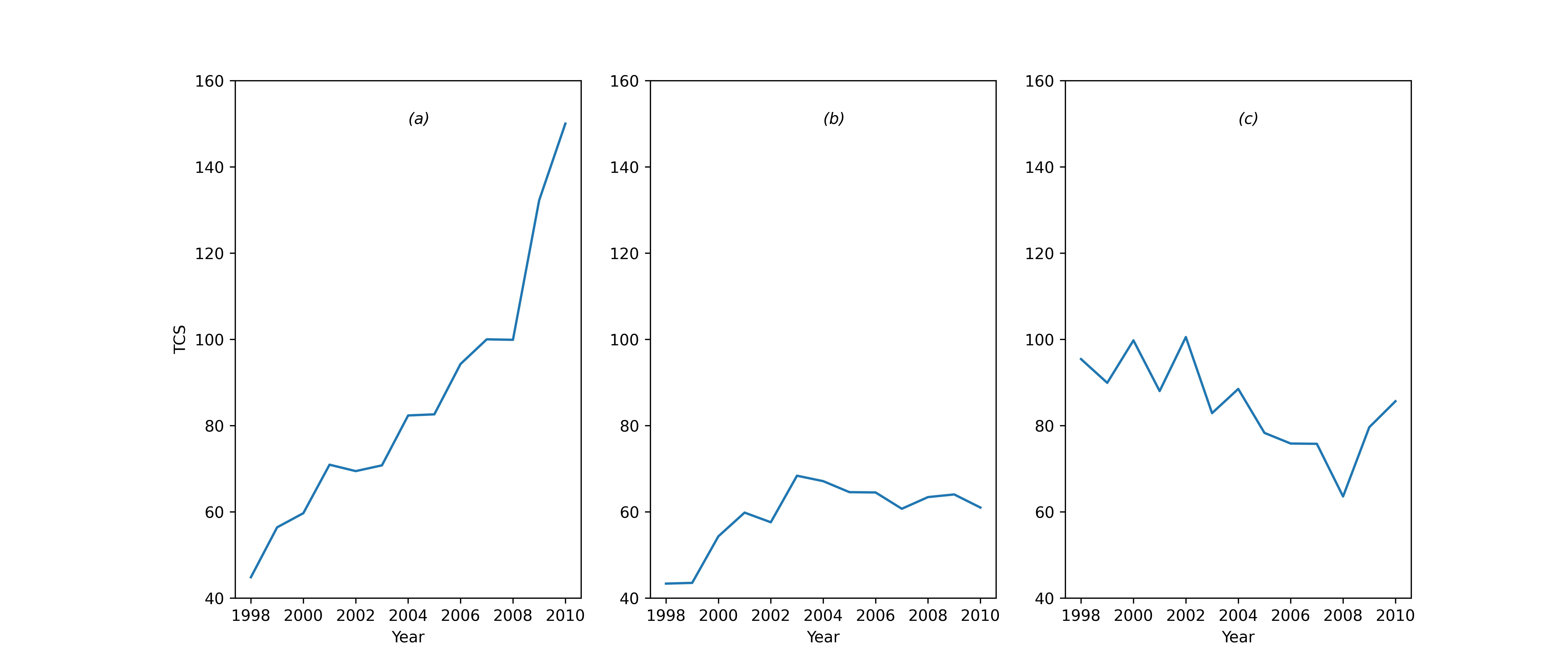}
\caption{Sample Topic Timeseries. Three different example topic time series for the 1998$-$2010 corpus appear in this figure which exemplify common behaviors seen in the population of topic timeseries.}
\label{fig:sample_topic_timeseries}
\end{figure*}

To determine the growth rate of research areas (Question 2) we calculate the time series of TCS for each topic \( t \). We may do this by breaking the corpus up into yearly groupings and then reapplying the TCS pipeline to derive TCS(\( t, y \)) where \( t \) is the research topic and \( y \) is the year for which TCS is calculated. Example time series appear in Figure \ref{fig:sample_topic_timeseries} calculated using the 1998$-$2010 corpus. Behavior in these plots is indicative of the larger population. Topic time series may clearly increase year on year (a), may exhibit an increase and then plateau at a certain level (b) or may simply remain at the same level or decrease overall. Significant variation from year to year may occur (c).

These time series may be analyzed to determine the compound annual growth rate of the Topic Contribution Score for a topic \( t \) or TCS\_CAGR(\( t \)). The equation for its calculation is
\begin{equation}
    \text{TCS\_CAGR}(t) = \left[ \frac{\text{TCS}(t, y_f)}{\text{TCS}(t, y_0)} \right]^{\frac{1}{P}} - 1,
\end{equation}
where TCS(\( t, y_f \)) is the TCS for topic \( t \) measured in the final year of the time series, TCS(\( t, y_0 \)) is TCS measured for the first year in the time series and \( P \) is the number of years in the time series. The observed variability (Figure \ref{fig:sample_topic_timeseries}c) presents the concern that the derived TCS\_CAGR values may fail to adequately represent the growth rate (or lack thereof) of TCS. To understand if this is the case or not we have also examined using a weighted TCS\_CAGR metric in which the first and last two years of TCS(\( t, y \)) were averaged. This did not appreciably change the resulting distribution of TCS\_CAGR values for the population of all topics.

The overall “importance” of research topics is a key question (Question 3). In this work, we associate importance with how engaged the community is in a given topic. We consider this engagement, or interest, to be the combination of how much a topic is studied by researchers as well as how much the research community for this topic is growing (or not). The most interesting topics are those topics which are both growing rapidly and have significant, but still growing, communities. We have tried to parameterize this interest using both the previously defined metrics of TCS and TCS\_CAGR. The “Research Interest” (RI); that is, how much overall interest the research community places on a given topic, is a gauge of its research engagement by the community in that topic, which may be quantified from these measures via
\begin{equation}
    \text{RI}(t) = [\text{TCS\_CAGR}(t) + O] * \text{TCS}(t),
\end{equation}
where \( t \) is the given topic, RI(\( t \)) is the Research Interest for topic \( t \), TCS(\( t \)) is the Topic Contribution Score for topic \( t \) and TCS\_CAGR(\( t \)) is the Topic Contribution Compound Annual Growth Rate of topic \( t \) and \( O \) is a constant offset to avoid the growth rate value being negative. Based on the data in this paper we utilize an offset of 0.05 (which will bring the lowest value of TCS\_CAGR in our ten run dataset just above zero).

We come now to the final question which is to understand whether this approach adequately models human perception of which areas of research are important (Question 4). To investigate this we utilized the Science Frontier panel chapters 1 - 4 for the 2010 National Academies’ Astronomy and Astrophysics Decadal Survey (“DS2010” corpus) which identify decadal priorities for the astronomy community. We split this chapter material into individual paragraphs, dropped content related to the formatting of the page (the page header), and applied the other cleaning criteria as before for the literature. There were 1020 resulting paragraphs available for comparison after this filtering. We used the same code as before to extract features from the DS2010 corpus paragraphs and applied the 1998 - 2010 literature topic models. Using these models we derived a topic contribution score by topic (TCS\(_{DS2010} (t, d)\)) for each paragraph (x) in the 2010 Survey.

The DS2010 TCS was derived by topic by calculating and summing over the derived inference for topics in all remaining paragraphs as given by
\begin{equation}
    \text{TCS}_{DS2010}(t) = \sum_{d=P_1}^{d=P_n} I_{DS2010}(t, d),
\end{equation}
where \( t \) is a given topic, \( d \) is the document (science frontiers paragraph), and \( I_{DS2010}(t, d) \) is the LDA inference score for topic \( t \) in paragraph \( d \) in the DS2010 corpus.

We may do something similar for whitepapers which have been submitted to the 2010 Decadal Survey. We utilized 274 submitted whitepapers by extracting and combining the whole text of the title, any abstract (not always present) and the main body of the whitepaper text. References were stripped out and unused. TCS\(_{whitepapers}\) may be defined as
\begin{equation}
    \text{TCS}_{whitepapers}(t) = \sum_{x=X_1}^{x=X_n} I_{whitepapers}(t, x),
\end{equation}
where \( t \) is a given topic, \( x \) is the ID of the text block representing the above indicated contents of the whitepaper (title, abstract, main body and excepting references) and \( I_{whitepapers}(t, x) \) is the LDA inference score for topic \( t \) in whitepaper text block \( x \) from a collection of size n.

We investigated the association between the Decadal Survey metrics TCS\(_{DS2010}\) and TCS\(_{whitepapers}\) to the literature-based metrics of TCS\(_{1998-2010}\), TCS\_CAGR\(_{1998-2010}\), and RI\(_{1998-2010}\). We created topic models for ten runs, each run initialized with a different random seed and producing 125 topic models. We then tested by cross-comparing Decadal Survey- and literature-based metrics to determine the value of any correlation between them in the data of each run. Using these results we then calculated the mean correlation values and standard errors of each test. These mean correlation results appear in Table \ref{tab:mean_results}. One post-processing filtering step was taken before calculation of the metrics; we removed one significant recurring non-astronomy topic, “Gravity Waves”, from each of the ten runs which left 124 topics for analysis.

\begin{table*}[htbp]
\centering
\label{tab:mean_results}
\begin{tabular}{l c c c c}
 \hline
 \textbf{Literature Metric} & \multicolumn{2}{c}{\textbf{Mean Correlation Coefficient}} & \multicolumn{2}{c}{\textbf{Mean P-value}} \\ 
 & \textbf{Pearson} & \textbf{Spearman} & \textbf{Pearson} & \textbf{Spearman} \\
 \hline
 \multicolumn{5}{c}{\textit{Literature Metric vs TCS\(_{DS2010}\)}} \\
 % & & & & \\
 TCS\(_{1998-2010}\) & 0.40 $\pm$ 0.2 & 0.51 $\pm$ 0.03 & 0.00014 & 8.6e-7 \\
 TCS\_CAGR\(_{1998-2010}\) & 0.47 $\pm$ 0.02 & 0.30 $\pm$ 0.02 & 5.1e-7 & 0.009 \\
 RI\(_{1998-2010}\) & 0.62 $\pm$ 0.03 & 0.57 $\pm$ 0.02 & 3.7e-9 & 8.5e-9 \\
 & & & & \\
 \multicolumn{5}{c}{\textit{Literature Metric vs TCS\(_{whitepapers}\)}}\\
 % & & & & \\
 TCS\(_{1998-2010}\) & 0.51 $\pm$ 0.2 & 0.61 $\pm$ 0.02 & 4.7e-8 & 1.3e-9 \\
 TCS\_CAGR\(_{1998-2010}\) & 0.43 $\pm$ 0.1 & 0.31 $\pm$ 0.02 & 5.4e-6 & 0.003 \\
 RI\(_{1998-2010}\) & 0.68 $\pm$ 0.2 & 0.65 $\pm$ 0.02 & 1.5e-13 & 4.0e-12 \\
 \hline
\end{tabular}
\caption{Mean results of tests of Pearson and Spearman correlation for Decadal Survey documents using ten runs of our LDA model pipeline. TCS\(_{DS2010}\) and TCS\(_{whitepapers}\) cross-correlation tests are shown for various metrics derived from the 1998$-$2010 astronomical literature. In most cases, the Research Interest metric RI\(_{1998-2010}\) outperforms or equals the other literature-based metrics. Errors are the standard error.}
\end{table*}

\begin{figure}[t!]
\centering
\includegraphics[width=.8\linewidth]{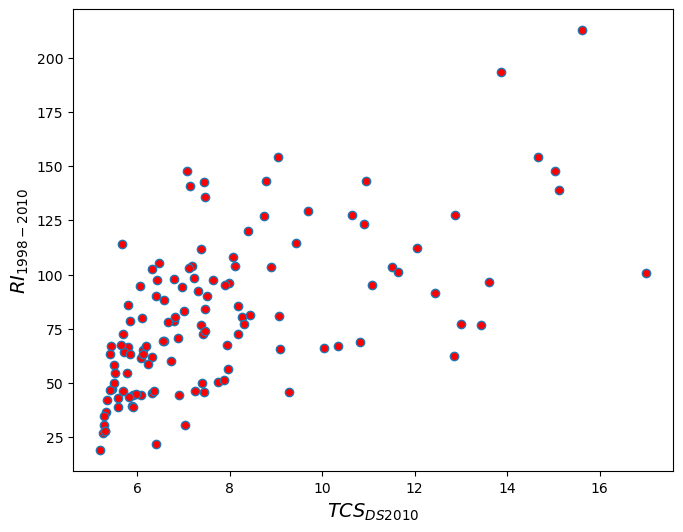}
\caption{Literature Research Interest compared to TCS for Decadal Survey content. An example plot of one run comparing the 1998 - 2010 literature metric RI\(_{1998-2010}\) versus the 2010 Decadal Survey Topic Contribution Score (TCS\(_{DS2010}\)) by topic. Each red dot represents a topic. The data indicate a significant, but weak-to-moderate, correlation exists. The mean Spearman correlation for ten runs is \( R_{mean} = 0.57 \pm 0.02 \). Estimated errors are of the same size or smaller than the symbols.}
\label{fig:literature_research_interest_tcsds2010}
\end{figure}

\begin{figure}[t!]
\centering
\includegraphics[width=.8\linewidth]{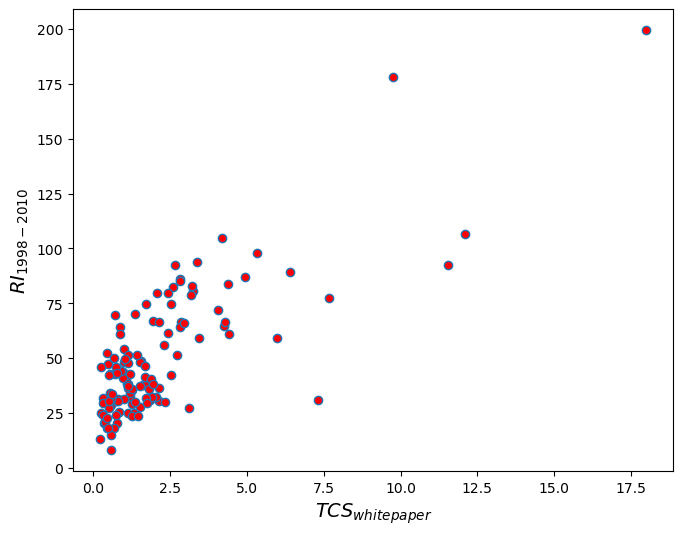}
\caption{Literature Research Interest compared to TCS for the Decadal Survey whitepapers. An example plot of one run comparing the 1998 - 2010 literature RI versus the 2010 decadal submitted whitepaper Topic Contribution Score (TCS\(_{whitepaper}\)) by topic. Each red dot represents a topic. The data indicate a significant, but moderate, Spearman correlation exists (\( R_{mean} = 0.65 \pm 0.02 \)). Estimated errors are of the same size or smaller than symbols.}
\label{fig:literature_research_interest_tcswhitepaper}
\end{figure}

Because a priori we have no reason to believe the relationships are necessarily linear, we performed correlation tests using both Pearson and Spearman statistics. As may be seen in Table \ref{tab:mean_results}, the RI\(_{1998-2010}\) metric generally showed the strongest correlation of the literature metrics, the only exception being for the Spearman-based tests of the Decadal Survey whitepapers where both RI\(_{1998-2010}\) and TCS\(_{1998-2010}\) are statistically equivalent. Figures \ref{fig:literature_research_interest_tcsds2010} and \ref{fig:literature_research_interest_tcswhitepaper} respectively show example results from one run for the literature metric RI\(_{1998-2010}\) compared to TCS\(_{DS2010}\) or TCS\(_{whitepapers}\). In both of these plots errors have been estimated using bootstrap error estimation and the errors are of the same size or smaller than the symbols in these figures.

As a check on how our metrics may relate to human perception of the importance of the research we investigated a different metric to compare to our TCS-based ones (e.g. TCS\(_{DS2010}\), TCS\(_{whitepapers}\), TCS\(_{1998-2010}\), TCS\_CAGR\(_{1998-2010}\), and RI\(_{1998-2010}\)). Citations of a paper are a known metric for assessing research impact (\cite{garfield1973citation}) and we have attempted to create a citation rate metric for each topic based on the papers associated with a given topic. One challenge for development of this metric is that the ensemble of associated papers are published at different times and therefore the measured citation rate for each paper needs to be normalized to some common interval of time. We have attempted to estimate the lifetime citation rate of each paper and then take the mean of each paper associated with a topic in order to develop a topic Mean Lifetime Citation Rate (MLCR, see Appendix B).

\begin{table*}[htbp]
\centering
\label{tab:comparison_metrics}
\begin{tabular}{l c c c c}
 \hline
 \textbf{Compared Metric} & \multicolumn{2}{c}{\textbf{Mean Correlation Coefficient}} & \multicolumn{2}{c}{\textbf{Mean P-value}} \\ 
 & \textbf{Pearson} & \textbf{Spearman} & \textbf{Pearson} & \textbf{Spearman} \\
 \hline
 \multicolumn{5}{c}{\textit{Literature-based Metrics}}\\
 TCS\(_{1998-2010}\) & 0.08 $\pm$ 0.2 & 0.08 $\pm$ 0.03 & 0.5 & 0.32 \\
 \textbf{TCS\_CAGR\(_{1998-2010}\)} & \textbf{0.75 $\pm$ 0.01} & \textbf{0.70 $\pm$ 0.01} & \textbf{6.6e-21} & \textbf{1.8e-17} \\
 RI\(_{1998-2010}\) & 0.42 $\pm$ 0.2 & 0.46 $\pm$ 0.02 & 2.3e-5 & 8.4e-6 \\
 & & & & \\
 \multicolumn{5}{c}{\textit{Decadal Survey-based Metrics}} \\
 TCS\(_{DS2010}\) & 0.51 $\pm$ 0.2 & 0.43 $\pm$ 0.02 & 3.7e-7 & 3.7e-5 \\
 TCS\(_{whitepaper}\) & 0.43 $\pm$ 0.1 & 0.32 $\pm$ 0.02 & 1.1e-5 & 0.01 \\
 \hline
\end{tabular}
\caption{Mean results of tests of the Pearson and Spearman statistics for various TCS-based metrics compared to MLCR (\( I_{paper} \geq 0.2 \)) for ten runs. The topic growth rate for the literature, TCS\_CAGR\(_{1998-2010}\) (in bold font), provided the strongest correlation. TCS\(_{1998-2010}\) is uncorrelated. Errors are the standard error.}
\end{table*}

\begin{figure}[t!]
\centering
\includegraphics[width=.8\linewidth]{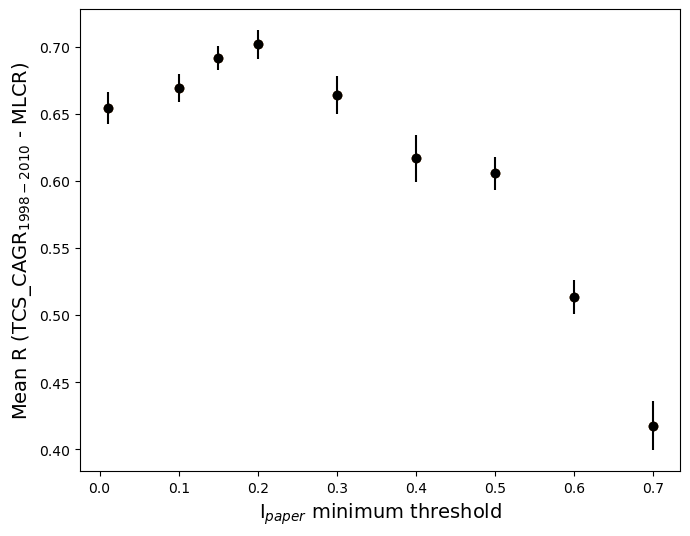}
\caption{Comparison of minimum topic inference thresholds to the mean TCS\_CAGR\(_{1998-2010}\) versus MLCR Spearman correlation. The higher the value of the threshold the greater the paper content (used to calculate MLCR) must be associated with a given topic. The correlation is maximized at a minimum paper inference threshold of \( I_{paper} \sim 0.2 \). Mean TCS\_CAGR\(_{1998-2010}\) and MLCR were calculated from the data from ten runs and errors are standard errors. The leftmost point is for a minimum paper threshold of \( I_{paper} = 0.01 \).}
\label{fig:comparison_minimum_topic_inference}
\end{figure}

In calculating the topic MLCR we needed to apply some filtering. First, we filtered out papers with a very low number of citations (less than 10 citations). Next we needed to determine which papers were most closely related to a given topic. This is critical because if we do not apply some filter then all papers in our corpus are associated with all topics and we will have the same calculated value of MLCR for each topic. There is no a priori obvious criteria for association of a paper (d) to a topic so we investigated applying different minimum inference thresholds to filter out the papers which contained less content than a given threshold for topic \( t \), as gauged by the paper topic inference, or \( I_{paper}(t, d) \), before calculation of MLCR. We ran a grid of tests (all metrics to MLCR in each of ten runs generating models with different random seeds) to check for the strength of the mean correlation between TCS-based metrics and the various MLCR at different paper minimum inference thresholds. A minimum paper inference threshold value of ~ 0.2 was found to yield the best mean correlation in all cases where a TCS-based metric was correlated. See Figure \ref{fig:comparison_minimum_topic_inference} for an example plot of these tests for the mean TCS\_CAGR\(_{1998-2010}\) metric. A final filtering step was to drop any topic which had fewer than three papers associated with it (this was rare at low \( I_{paper} \) minimum thresholds but occurred with increased frequency for topics as the minimum paper threshold for association with a topic was increased).

\begin{figure}[t!]
\centering
\includegraphics[width=.8\linewidth]{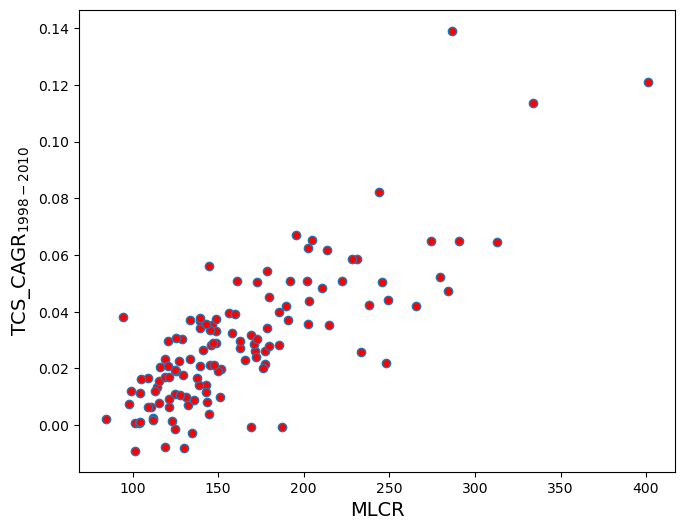}
\caption{Comparison of literature topic growth to literature citation rate. Example results of one run comparing TCS\_CAGR\(_{1998-2010}\) and to the estimated topic MLCR. Each red dot represents a topic. The mean TCS\_CAGR\(_{1998-2010}\) is significantly correlated (\( R_{mean} = 0.75 \pm 0.01 \); standard error estimated from 10 runs). In this plot the individual topic errors for TCS\_CAGR\(_{1998-2010}\) are smaller than the symbol size and MLCR errors were not estimated. We filtered papers with low inference values (\( I_{paper} < 0.2 \)) before calculating MLCR for each topic (see text for an explanation).}
\label{fig:comparison_growth_citation}
\end{figure}

Comparing results of TCS-based metrics determined from the ten runs of our pipeline (see Table \ref{tab:comparison_metrics}) we consistently find TCS\_CAGR\(_{1998-2010}\) is best correlated to MLCR regardless of the choice of statistic (see Figure \ref{fig:comparison_growth_citation} for an example run). Surprisingly TCS\(_{1998-2010}\) is not significantly correlated with MLCR and RI\(_{1998-2010}\) is not a useful metric to compare to MLCR because of the uncorrelated contribution of TCS\(_{1998-2010}\) and should be discarded. When we compare Decadal Survey-based metrics with MLCR we find that both TCS\(_{DS2010}\) is significantly correlated with MLCR although the correlation is not as strong as for TCS\_CAGR\(_{1998-2010}\). TCS\(_{whitepaper}\) is not significantly correlated when testing using the Spearman statistic. In the case of significant correlations the comparison of Pearson and Spearman statistics indicates lower overall comparable correlation coefficients for Spearman. We interpret this as largely due to some significant outliers in the dataset driving the Pearson correlation coefficient slightly higher (see Figure \ref{fig:comparison_growth_citation} for an example which does show a ‘tail’ away from the main body of points for higher values of MLCR). Spearman is likely the best choice of statistic for quantifying the relationship with MLCR.

Public repositories containing code and data used by this work include:
\begin{itemize}
    \item Analysis Software: \cite{brian_thomas_2024_10846255}
    %\url{https://github.com/brianthomas/ml_strat_prioritization}
    \item Topic Modeling: \cite{buonomo2024code}
    %\url{https://github.com/abuonomo/topic-emergence-ADS}
    \item LDA Training Dataset: \cite{thomas2024dataset}
\end{itemize}

\section*{Discussion}

Our results indicate it is possible to develop metrics based on LDA modeling of the research literature which may be used to quantify how much activity or ‘interest’ there is in certain research topics (TCS) and how much that interest is changing (TCS\_CAGR). Across ten runs which regenerated the LDA models based on different random seeds we see consistent behavior where the best correlation between either the Decadal Survey frontier panel report content (TCS\(_{DS2010}\)) or submitted whitepaper content (TCS\(_{whitepaper}\)) and that of the astronomical literature (1998 - 2010) is found by the Research Interest metric (RI\(_{1998-2010}\)) which folds together the behavior of the other metrics.

Interestingly, we find only TCS\_CAGR\(_{1998-2010}\) and not RI\(_{1998-2010}\) to be significantly correlated with the Mean Lifetime Citation Rate of topics (MLCR). This result may indicate that the Decadal Survey and the submitted whitepapers place significant emphasis on popular established research (as captured by the TCS metric) and both the Decadal Survey and whitepapers may under-emphasize new and growing research topic areas (as captured by TCS\_CAGR). The very significant moderate correlation between MLCR and TCS\_CAGR\(_{1998-2010}\) is also an independent confirmation that our approach can tease out information on which research areas (topics) are important.

We note that in all cases our correlations, although significant, are of only moderate strength and the resultant coefficient of determination (\( R^2 \)), a measure of how much of the variability in one variable results from variation in the other, is moderately weak (mean \( R^2 \approx 0.4 \) for the RI\(_{1998-2010}\) to TCS\(_{DS2010}\) correlation, for example). At least several reasons may explain why. First, in some cases we may be undersampling the trend in the topic time series as we are selecting only a restricted set of astronomy journals which would lead to some variation in measured TCS\_CAGR. Another issue is that our technique models the language present in scientific abstracts, but this may be significantly different from the language present in the 2010 Decadal Survey corpus and could thereby result in lowering the TCS\(_{DS2010}\) values. Finally, while nearly all topics are relevant to the Decadal Survey, there remains for each LDA run an admixture of topics which are not germane to the Decadal Survey such as “solar wind” and “coronal mass ejection,” so the mapping of research topic matter is not one-to-one between the models and the 2010 Decadal Survey and these non-Survey topics will lower the correlation we derive. Removing these topics, however, is not “clean,” as we can expect there is no simple way to isolate heliophysics from astrophysics topics as many of their research areas overlap somewhat. Culling to use simply “astronomy journals”' is not a perfect solution either; in practice astronomy journals will accept and publish according to their interest, not according to our desired categories for astronomy.

A word should be mentioned about the nature of the topics. In general these appear to be fairly specific topics which collect papers in a helpful manner such as “monte carlo simulation and black hole mass” or “mass accretion, accretion rate, and young stellar object”. Nevertheless, in each run we still derive a small number of less-specific topics which indicate broad research areas which collect papers from across various areas of astronomy. A good example of this is the topic “magnetic field, field strength, field line” or “soft x-ray, x-ray emission, x-ray light curve”. Presumably moving to a higher number of topics when modeling would help to better divide these papers into separate topics, but doing so would lose coherence in the modeling producing poorer models. We simply do not have sufficient data at this time to further increase the number of topics. Another concern is the potential presence of “semantic drift,” whereby the description of a field changes appreciably over the time period being studied. Again because of the limited amount of data we had available, we were unable to investigate this and assumed little to no change in meaning for our topics over the roughly ten years of research we examined (e.g., the “star formation” research topic meant more or less the same thing across the years of 1998 to 2010).

\begin{figure}[t!]
\centering
\includegraphics[width=.8\linewidth]{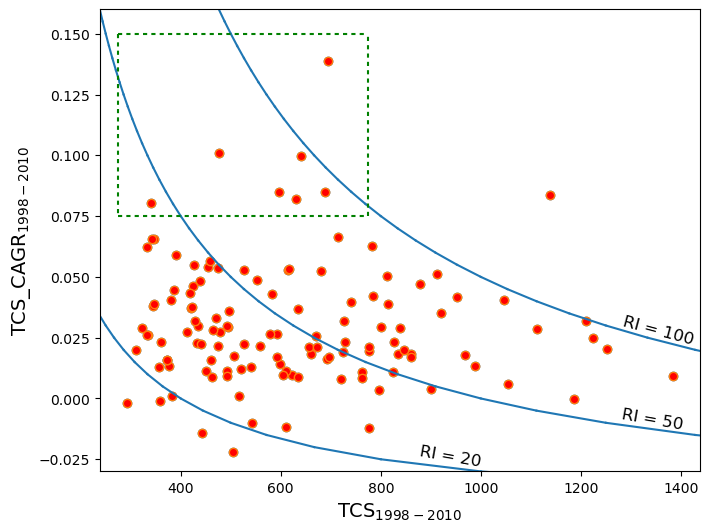}
\caption{Example use of literature-based metrics TCS and TCS\_CAGR. Blue lines represent constant RI (where \( O = 0.05 \)). Extremes of the plot distribution may be of some utility to planners. The green dashed box, for example, marks topics which are not as popular as many others (TCS \( \lesssim 700 \)) but which are exhibiting very high growth in interest. Conversely, an examination of low RI, and/or low TCS\_CAGR, topics may reveal areas of lessening strategic importance. Data are from one run using the 1998 to 2010 astronomical corpus.}
\label{fig:tcs_vs_tcscagr}
\end{figure}

Nevertheless, despite these issues there is still immediate value in applying these metrics to the science prioritization process. Unlike the MLCR, a metric based on citation rates, which is a lagging measure of topic research impact, these new measures are leading indicators, which makes them attractive for planning. As shown in Figure \ref{fig:tcs_vs_tcscagr}, one potential application may be to utilize the TCS\_CAGR measure to identify probable future impactful research topics and papers, thus creating valuable curated reading for participants in the Decadal Survey process. Another application may be to utilize a TCS to TCS\_CAGR scatterplot (see Figure \ref{fig:tcs_vs_tcscagr}) for scientific papers to investigate the topics which fall at the extremes of the topic distribution and thereby uncover potentially new interesting areas for investment, or conversely, discover areas of investment which may need to be re-evaluated and descoped.

This is exploratory work and looking forward there are many possibilities for further research. We expect that this experiment can and should be conducted with a more recent research corpus to continue to qualify the relationship between TCS\_CAGR and citation rates/research impact. It would also be fruitful to apply this work in other science domains which have Decadal Surveys or similar studies to use as a reference for how that community is prioritizing its science investments in order to see if the relationship with RI holds for those documents. Finally, we note that the ML technology employed in this study is now at least a generation old! Leveraging newer, more advanced NLP techniques and language models which are trained for scientific data such as AstroBERT (\cite{grezes2022improving}) or perhaps more powerful, but general, large language models based on GPT-4 (\cite{openai2023gpt4}) may help us to better quantify and pull useful features from the scientific literature and result in improved predictions.

\section*{Summary}

We have found there is potential to utilize Machine Learning (ML) to aid in science planning and prioritization. Using open-source ML algorithms we have explored different derived metrics that may be used to predict future research interest, as well as research topics that may be of declining future research interest. We trained topic models using a corpus of scientific papers drawn from high-impact journals and then tested the performance of our derived metrics using strategic publications which project what the field of astronomy should be doing in the coming decade. These reference materials include the science frontier panel chapters of the 2010 National Academies’ Decadal Survey report for Astronomy and Astrophysics (“DS2010”) and the submitted Decadal Survey whitepapers (“whitepapers”). In both cases we find some agreement between the predictions of our metrics for the astronomical literature (1998 - 2010) and the predictions of these metrics when applied to either the DS2010 or whitepaper materials. After examination of all tests, we find that the best performing metric was one which blended the current popularity or interest in a topic area (“TCS”) with a measure of its growth (“TCS\_CAGR”). This metric, “Research Interest” (RI), shows a very significant moderate correlation (\( R \approx 0.6 - 0.7 \)) between the astronomical literature and with these Decadal Survey-based materials.

We also checked for independent confirmation of the performance of our metrics. Using a citation rate-based measure for the topics, “Mean Lifetime Citation Rate” (MLCR), we found a significant moderate correlation (\( R \approx 0.7 \)) to the topic growth rate TCS\_CAGR exists. Surprisingly TCS was not significantly correlated with MLCR. We interpret this to possibly mean that the Decadal Survey materials under-emphasize topic areas which are growing – which conversely is the best predictor of whether research will have future impact.

While this is exploratory research there may be some immediate application of this work. A process such as ours can easily produce a useful curated list of published work for future Decadal Survey participants or might be applied by science planners to uncover potential research areas for increased (or decreased) investment.

\section*{Declaration of Generative AI and AI-assisted technologies in the writing process}
During the preparation of this work the authors used ChatGPT in order to integrate the draft text with the Astronomy and Computing Latex template. After using this tool/service, the authors reviewed and edited the content as needed and take full responsibility for the content of the publication.

\section*{Acknowledgements}
The authors acknowledge the support of NASA grant NNH22ZDA001N-HTM, an internal grant from the NASA Science Mission Directorate Chief Technologist, and support of the NASA Office of the Chief Scientist. The authors also acknowledge the use of the HelioCloud environment (https://heliocloud.org) where the processing and analysis were performed.

%% The Appendices part is started with the command \appendix;
%% appendix sections are then done as normal sections
% \appendix

% \section{Appendix title 1}
% %% \label{}

% \section{Appendix title 2}
%% \label{}

\appendix

\section{Top High-Impact Astronomy and Astrophysics Journals 1999-2010}

We extracted journal data from the SciMago website for the Astronomy and Astrophysics category (\url{https://www.scimagojr.com/journalrank.php}). This site provides the SciMago Journal Rank Indicator (SJRI; see SCImago Journal \& Country Rank (2021), [26]), which we averaged and then ranked for all journals for the time period 1999 to 2010 (1998 data were not available). Journals which are not unambiguously dedicated to astronomy were filtered out (for example “Annual Review of Earth and Planetary Sciences”, “Solar Physics”, and “Icarus”) and journals were selected that had an average $SJRI \geq 1.0$. The resulting list is shown in Table \ref{tab:high_impact_journals}.

\begin{table*}[t!]
\centering
\label{tab:high_impact_journals}
\begin{tabularx}{\textwidth}{|X|c|}
 \hline
 \textbf{Journal} & \textbf{Mean SJRI (1999-2010)} \\ 
 \hline
 Annual Review of Astronomy and Astrophysics & 14.1 \\
 Astrophysical Journal, Supplement Series & 6.9 \\
 Astrophysical Journal Letters & 5.4 \\
 Astronomy and Astrophysics Review & 4.5 \\
 Astronomical Journal & 4.1 \\
 Monthly Notices of the Royal Astronomical Society & 3.5 \\
 Astrophysical Journal & 3.2 \\
 Astronomy and Astrophysics & 2.7 \\
 Publications of the Astronomical Society of the Pacific & 2.4 \\
 Acta Astronomica & 2.0 \\
 Astroparticle Physics & 2.0 \\
 Publication of the Astronomical Society of Japan & 1.8 \\
 New Astronomy & 1.6 \\
 Publications of the Astronomical Society of Australia & 1.0 \\
 \hline
\end{tabularx}
\caption{Top High-Impact Astronomy Journals 1999 to 2010.}
\end{table*}

\section{Calculation of the Estimated Mean Lifetime Citation Rate}

\begin{figure*}
    \centering
    \includegraphics[width=\textwidth]{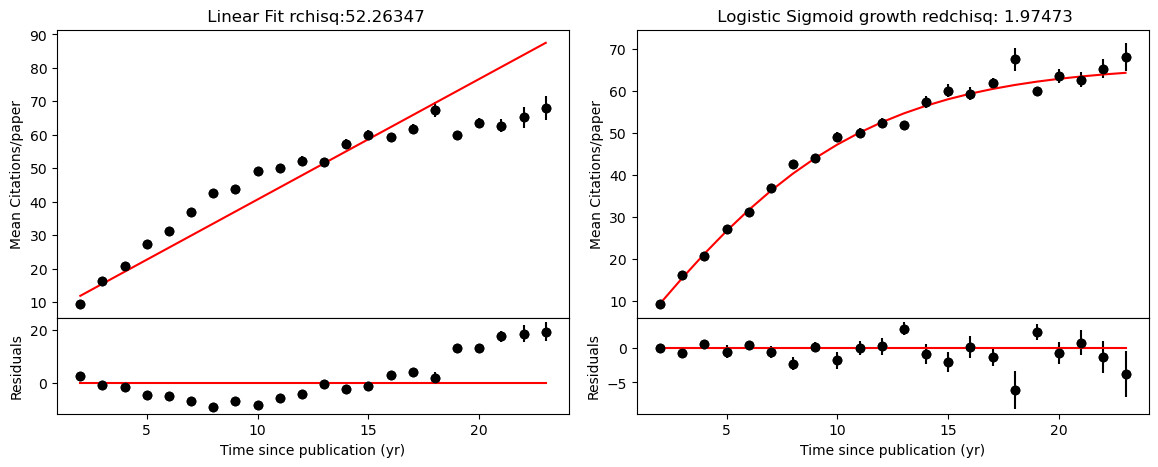}
    \caption{Comparison of a linear and a logistic sigmoid function (see Equation~\ref{eq:B1}). Fits to the mean citation data by year for high-impact journal papers (1998 to 2019). The logistic sigmoid function fits the data significantly better than the linear function. Mean citation uncertainties were calculated using bootstrap error estimation of the standard error.}
    \label{fig:sigmoid_fit}
\end{figure*}

The estimated Mean Lifetime Citation Rate (MLCR) is calculated by first fitting data drawn from 1998 to 2019 high-impact astronomy journal papers mean citation data with a logistic sigmoid function, which is given by

\begin{equation}
    \text{Estimated Citations} (t) = \frac{\beta}{1 + e^{-t/\alpha}} + \gamma, \label{eq:B1}
\end{equation}

where \( t \) is the time since publication (in years), \( \alpha \) is the timescale of change constant, \( \gamma \) is a bias offset to adjust mean starting citations and \( \beta \) is a scaling constant. Uncertainties in mean citations per paper were estimated using bootstrap error estimation (standard error for the mean). Figure \ref{fig:sigmoid_fit} shows the results of fitting both linear and logistic sigmoid functions to the mean paper citations. The logistic sigmoid function significantly performs better than the linear function and yields \( \beta = 138.7 \pm 2.2 \) mean citations/paper, \( \alpha = 5.6 \pm 0.2 \) years and a bias offset \( \gamma = -72.5 \pm 2.2 \) mean citations/paper. 

Using this relationship, we obtained the estimated Lifetime Citations (LC) for each of the papers (assuming that the fit parameters \( \beta \), \( \alpha \), and \( \gamma \) were the same for all papers) with the following calculation:

\begin{equation}
    \text{Lifetime Citations} (p) = \frac{\text{Cites}(p) - \gamma}{1 + e^{-t(p)/\alpha}}, \label{eq:B2}
\end{equation}

where \( p \) is the given paper, \( t(p) \) is the year paper \( p \) was published and \(\text{Cites}(p)\) is the number of papers citing paper \( p \), as measured in 2021, the year we extracted the ADS citation data. The estimated Mean Lifetime Citation Rate (MLCR) for any topic is then the mean LC for all papers associated with the topic.

%% If you have bibdatabase file and want bibtex to generate the
%% bibitems, please use
%%
\bibliography{ml_strat_paper}
\bibliographystyle{elsarticle-harv} 

%% else use the following coding to input the bibitems directly in the
%% TeX file.

%%\begin{thebibliography}{00}

%% \bibitem[Author(year)]{label}
%% For example:

%% \bibitem[Aladro et al.(2015)]{Aladro15} Aladro, R., Martín, S., Riquelme, D., et al. 2015, \aas, 579, A101

%%\end{thebibliography}

\end{document}